\documentclass[aps,pre,twocolumn,showpacs,superscriptaddress,preprintnumbers,ams
math,amssymb]{revtex4}

\usepackage{amsmath}
\usepackage{graphicx}
\usepackage{dcolumn}
\usepackage{bm}
\usepackage{float}

\begin{document}

\author{M. E. Dieckmann}
\affiliation{Department of Science and Technology, Link\"oping University, SE-60174 Norrk\"oping, Sweden}

\author{G. Sarri}\affiliation{School of Mathematics and Physics, Queen's University Belfast, BT7 1NN, UK}

\author{D. Doria}\affiliation{School of Mathematics and Physics, Queen's University Belfast, BT7 1NN, UK}

\author{A. Ynnerman}
\affiliation{Department of Science and Technology, Link\"oping University, SE-60174 Norrk\"oping, Sweden}

\author{M. Borghesi}\affiliation{School of Mathematics and Physics, Queen's University Belfast, BT7 1NN, UK}

\date{\today}

\title{Particle-in-cell simulation study of a lower-hybrid shock}

\begin{abstract}
The expansion of a magnetized high-pressure plasma into a low-pressure ambient medium is examined with particle-in-cell (PIC) simulations. The magnetic field points perpendicularly to the plasma's expansion direction and binary collisions between particles are absent. The expanding plasma steepens into a quasi-electrostatic shock that is sustained by the lower-hybrid (LH) wave. The ambipolar electric field points in the expansion direction and it induces together with the background magnetic field a fast E cross B drift of electrons. The drifting electrons modify the background magnetic field, resulting in its pile-up by the LH shock. The magnetic pressure gradient force accelerates the ambient ions ahead of the LH shock, reducing the relative velocity between the ambient plasma and the LH shock to about the phase speed of the shocked LH wave, transforming the LH shock into a nonlinear LH wave. The oscillations of the electrostatic potential have a larger amplitude and wavelength in the magnetized plasma than in an unmagnetized one with otherwise identical conditions. The energy loss to the drifting electrons leads to a noticable slowdown of the LH shock compared to that in an unmagnetized plasma.
\end{abstract}

\maketitle
\section{Introduction}


Collisions between unmagnetized clouds of electrons and ions at a speed in excess of the ion acoustic speed can trigger the formation of electrostatic shocks. An electrostatic shock is an ion acoustic wave that has steepened into a sharp density jump. Ion acoustic waves and electrostatic shocks are sustained by the following process.

Consider a plasma with a spatially varying number density. Thermal diffusion will let electrons flow from a region with a high plasma density towards one with a low density and their larger inertia implies that the ions can not follow them. The current, which arises from the electron redistribution, generates an electric field. The mean electric field is called the ambipolar electric field. Its amplitude value is such that the electrostatic force balances the electron's thermal pressure gradient force, which inhibits a net flow of electrons. 

The ambipolar electric field can accelerate ions from an interval with a high plasma density to one with a low plasma density. The ion redistribution alters the plasma density gradients and thus the ambipolar electric field. The changes of the ambipolar electric field and of the ion density distribution are out of phase and both oscillate around an equilibrium distribution in the form of ion acoustic waves. We obtain almost undamped ion density oscillations if the electrons are hotter than the ions.

The ion acoustic wave is the only wave mode in a nonrelativistic setting and in the absence of a background magnetic field, which can modulate the density of the ions. It is thus the only wave mode that can sustain an electrostatic shock \cite{Bardotti70,Forslund70,Forslund71,Karimabadi91,Sorasio06,Kato10,Sarri11,Eliasson14} in a collisionless unmagnetized plasma unless the collision speed is high enough to yield a partially magnetic shock \cite{Pohl12,Stockem14}. The density gradient at the electrostatic shock drives an ambipolar electric field, which puts the downstream region behind the shock on a higher positive potential than the upstream region ahead of it. This potential difference slows down the inflowing upstream plasma. The therefrom resulting compression of the plasma sustains the density gradient. Self-consistent and steady-state solutions of an ion acoustic wave that steepened into an electrostatic shock exist, provided that the speed of the upstream plasma measured in the shock frame does not exceed a few times the ion acoustic speed \cite{Eliasson14}. 

Electrostatic shocks are now routinely produced in laser-plasma experiments \cite{Dean71,Honzawa73,Bell88,Romagnani08, Morita10, Kuramitsu11, Ahmed13, Niemann13} and they attract considerable interest because they allow us to study in-situ some of the plasma processes that develop in remote astrophysical environments. An example are the shocks that ensheath the blast shells of supernova remnants \cite{Remington99}. 

A magnetized plasma supports several compressional wave modes and, hence, different types of shocks. Magnetohydrodynamic (MHD) shocks can be mediated by the Alfv\'en wave, by the slow magnetosonic mode, by the intermediate mode or by the fast magnetosonic mode. The particular mode is selected primarily by the angle between the shock normal and the magnetic field and the speed, with which the shock propagates. MHD shocks form on time scales that are longer than an inverse ion gyrofrequency. The thickness of their transition layers exceeds the gyroradius of the inflowing upstream ions in the shock's magnetic field. The best-known examples are probably the fast magnetosonic shock that forms between the solar wind and the Earth's magnetopause \cite{Sckopke83} and the solar wind termination shock \cite{Richardson08} that separates the heliosphere from the interstellar medium. 

A magnetized plasma does support more ion wave modes than the aforementioned MHD waves. In what follows we consider waves that travel orthogonally to the magnetic field. A perpendicular magnetic field of suitable strength will limit the electron mobility on spatial scales in excess of their gyroradius. An ion density gradient will nevertheless drive an ambipolar electric field because the electrons can still move on spatial scales below their gyroradius. The ambipolar electric field has a component that is antiparallel to the ion density gradient, which enforces an electron drift in the direction orthogonal to the magnetic field and to the density gradient. 

The electron response to the ambipolar electric field is altered by its gyro- and drift motion, which modifies in turn the dispersion relation of the electric field oscillations. This effect plays an important role at frequencies above the ion gyrofrequency. The MHD approximation breaks down at such high frequencies and it has to be replaced by a two-fluid approximation. The two-fluid approximation reveals the presence of the almost electrostatic wave branch, which is known as the lower-hybrid (LH) mode. A kinetic model reveals that the LH mode goes over into the ion cyclotron waves if its wavelength is no longer large compared to the ion's thermal gyroradius. The dispersion relation of LH waves is discussed in various approximations in Ref. \cite{Verdon09}.

LH oscillations can have a shorter wavelength and a higher frequency than magnetosonic waves. A LH wave can thus steepen into a shock faster and on a smaller spatial scale. A LH wave with a small wavelength is practically electrostatic and we may expect that a shock, which is mediated by this wave mode, is electrostatic too. So far the LH waves have received attention with respect to instabilities close to shocks \cite{McClements97}, but the observation of a LH shock per se has not yet been reported.

Collisions of magnetized plasma clouds have been widely examined by means of particle-in-cell (PIC) simulations, but the collision speeds were always too high to yield shocks that could be sustained solely by the electric cross-shock potential. The fast shocks in the aforementioned case studies formed primarily due to the magnetic rotation and reflection of the colliding plasmas \cite{Lembege92,Shimada00,Scholer03,Chapman05,Chapman14} and were caused to a lesser degree by the cross-shock electric field. The latter has been observed in situ at the Earth's perpendicular bow shock \cite{Walker04}, which is otherwise a fast magnetosonic shock. See Ref. \cite{Treumann09} for a recent review of magnetized nonrelativistic shocks.

Here we show by means of PIC simulations that LH shocks exist and we discuss how such a shock differs from electrostatic shocks in an unmagnetized plasma. The LH shock we observe is a transient structure like its counterpart in unmagnetized plasma. The lifetimes of a LH shock and of an electrostatic shock are, however, limited for different reasons. 

The narrow unipolar electric field pulse, which characterizes electrostatic shocks in unmagnetized plasma, is transformed into a broad shock transition layer by the ion acoustic turbulence, which is generated upstream of the shock by the shock-reflected ion beam \cite{Karimabadi91,Kato10,Dieckmann14}. 

The LH shock is modified by the magnetic field it is piling up as it expands into the upstream region. The magnetic pressure gradient force it exerts on the ambient ions in the upstream region is pre-accelerating them, which reduces the ion velocity change at the shock to a value that is comparable to or below the phase speed of the LH wave. The structure changes in time from a strong LH shock into what appears to be either a weaker shock or a nonlinear LH wave. A similar qualitative distinction of shocks and nonlinear waves based on the shock speed was given in Ref. \cite{Forslund71}. 

While the LH shock compressed the ambient plasma to the same density as the unmagnetized electrostatic shock, the ambient plasma that crosses the magnetized structure at late times is hardly compressed. This structure does, however reflect some of the incoming upstream ions, which is a signature of a collisionless shock. This nonlinear LH wave balances the ram pressure of the upstream plasma primarily with the gradient of the magnetic pressure and to a lesser degree with the thermal pressure of the downstream plasma. 

The expanding blast shell piles up the magnetic field ahead of the shock, thereby increasing the magnetic pressure in the ambient plasma. This suggests an the following long-term evolution of the plasma. Initially the ambient plasma is the upstream medium and the blast shell is the downstream medium that expands into the ambient plasma due to its thermal pressure. The gradual increase of the magnetic pressure gradient force suggests that in the long term the magnetic field may dominate the plasma dynamics. Eventually the ambient medium may obtain a large enough magnetic pressure to become the downstream region of a magnetosonic shock and the blast shell provides the fast upstream flow.  

This paper is structured as follows. Section 2 compares the solution of the linear dispersion of LH waves that propagate strictly perpendicularly to the background magnetic field with the noise spectrum that is computed by PIC simulation, exploiting the fact that the noise is strongest when its wave number and frequency match that of a plasma eigenmode \cite{Dieckmann04}. Section 3 shows by means of PIC simulations how a LH mode shock forms and how it changes with time into a nonlinear LH wave. The simulations reveal the electromagnetic signatures of LH shocks, which should be detectable in laser-generated plasma and which differ from those of the well-researched electrostatic shocks in unmagnetized plasma. We summarize our results in Section 4.


\section{Electrostatic waves in magnetized plasma}

We consider here the approximate solution of the linear dispersion relation of LH waves, which is based on a two-fluid approximation, that takes into account warm plasma effects and neglects electromagnetic effects. It is valid for large wavenumbers $k=|\mathbf{k}|$ and for waves that move strictly perpendicularly to the magnetic field $\mathbf{B}$ with amplitude $B_0$. The LH frequency 
\begin{equation}
\omega_{lh} = {\left [ {(\omega_{ce}\omega_{ci})}^{-1} + \omega_{pi}^{-2} \right ]}^{-1/2} 
\end{equation}
becomes a resonance frequency at low $k$, where thermal effects are negligible. The electron's thermal gyroradius $r_{ce} = v_{te} / \omega_{ce}$, where $v_{te}={(k_B T_e / m_e)}^{1/2}$ ($k_B, T_e, m_e$: Boltzmann constant, electron temperature and mass) is the electron thermal speed and $\omega_{ce} = eB_0 / m_e$ is the electron gyrofrequency. The ion's thermal gyroradius is $r_{ci}= v_{ti}/\omega_{ci}$. The plasma frequency of ions with the number density $n_{i0}$, the charge $q_i$ and the mass $m_i$ is $\omega_{pi} = {(q_i^2 n_{i0} / m_i \epsilon_0 )}^{1/2}$ and $\omega_{ci} = q_i B_0 / m_i$ is their gyrofrequency. The thermal speed of ions with the temperature $T_i$ is $v_{ti}={(k_B T_i / m_i)}^{1/2}$.


We consider wave vectors $\mathbf{k}$ with $\mathbf{k}\cdot \mathbf{B}=0$, a temperature ratio $T_e / T_i = 12.5$ and fully ionized nitrogen ions $N^{7+}$ with the number density $n_{i0} = 4 \times 10^{13}\textrm{cm}^{-3}$. The electron number density is $7n_{i0}$. The magnetic field strength is $B_0 =$ 0.85 T, yielding $\omega_{pi}=6\omega_{lh}$ and $\omega_{lh}=60\omega_{ci}$. The ion composition and number density are representative for the ambient plasma in laboratory experiments; more specifically, the ratio between the plasma frequencies of the electrons and ions and the gyrofrequencies are comparable to those used in Ref. \cite{Niemann13}. The magnetic field strength is selected such that the frequency of the LH branch at large wavenumbers becomes comparable to the ion plasma frequency. Magnetic field effects should develop fast enough to be detectable. The plasma $\beta \approx 0.2$ in the ambient medium implies that the magnetic pressure is high enough to balance the thermal pressure of the expanding dense medium.

We can neglect magnetic field effects on the ion motion if the wave frequencies are $\omega \approx \omega_{lh}$ and approximate the ion susceptibility as $\chi_i = \omega_{pi}^2 / (3v_{ti}^2k^2 - \omega^2)$. The electron susceptibility to LH oscillations is approximated as $\chi_e = \omega_{pe}^2/(v_{te}^2k^2 + \omega_{ce}^2)$. The dispersion relation is $1+\chi_i+\chi_e = 0$ or
\begin{equation}
\omega^2 = 3v_{ti}^2 k^2 + \frac{\omega_{pi}^2 (\omega_{ce}^2 + v_{te}^2 k^2)}{\omega_{pe}^2 + \omega_{ce}^2 + v_{te}^2 k^2}
\label{DispersionRelation}
\end{equation}

We compare the dispersion relation given by Eqn. \ref{DispersionRelation} to the electrostatic noise distribution, which is computed by the EPOCH PIC simulation code. The simulation resolves one spatial (x) direction and it has been initialized with the aforementioned plasma parameters. The simulation employs periodic boundary conditions and it resolves a box length of 16 mm or 96 $r_{ce}$ by 3200 grid cells. We run the simulation for $t_s$ = 4 ns, which resolves the fraction $\omega_{ci} t_s = 0.16$ of an ion gyro-orbit. The electron temperature is set to $T_e = 2$ keV.

Electrostatic waves are polarized along the simulation direction and they can be identified using the $E_x$ component. We obtain the noise distribution by sampling the electric field $E_x(x,t)$, by taking its Fourier transform over space and time and by squaring the modulus of the result. The power spectrum of the noise distribution of a PIC simulation peaks at frequencies, which are eigenmodes of the plasma and it can thus be used to reveal linearly undamped or weakly damped wave branches. 

\begin{figure}
\includegraphics[width=\columnwidth]{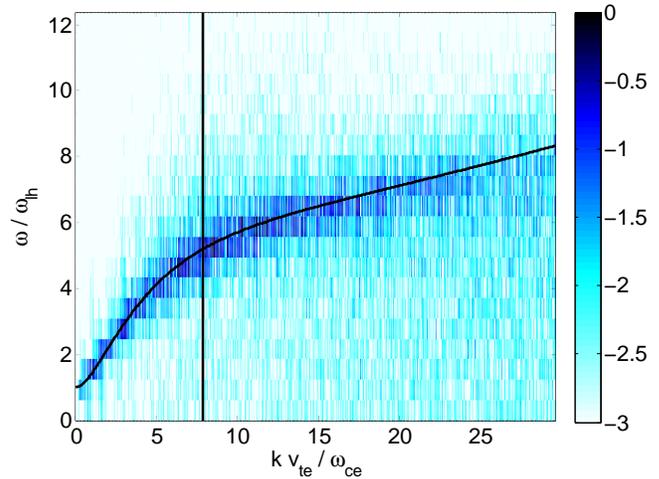}
\caption{The 10-logarithmic power spectrum ${|E_x(k,\omega)|}^2$ of the electrostatic noise, which has been computed by a PIC simulation. Overplotted is the solution of the linear dispersion relation. The vertical line corresponds to a wavelength of 0.1 mm for the considered plasma parameters.}\label{DispRel}
\end{figure}

Figure \ref{DispRel} shows the result. Strong noise that follows the solution of Eq. \ref{DispersionRelation} is observed up to a wave number $k r_{ce} \approx 20$. We thus identify this mode as the LH mode. 


We calculate the phase speed $v_{ph} \approx 2 \times 10^5$ m/s of the LH wave at the reference wavelength 0.1 mm using the noise distribution. We can compare the phase speed to the phase speed $c_s$ of the ion acoustic waves, which would be present if $B_0=0$. The ion acoustic speed $c_s = {(k_B (T_e + 3T_i)/m_i)}^{1/2}$ that takes into account a one-dimensional adiabatic expansion of ions and isothermal electrons amounts to $\approx 1.3 \times 10^5$ m/s for our plasma parameters giving $v_{ph} \approx 1.5c_s$. 

Electrostatic shocks, which are mediated by the ion acoustic wave, are routinely observed in laser-plasma experiments and we may assume that a magnetic field will not affect the expansion speed of the laser-generated blast shell on time scales of the order of an inverse LH frequency. LH shocks should form in the presence of a perpendicular magnetic field.

\section{PIC simulations of LH shocks in one and in two dimensions}

We perform PIC simulations with the aim to determine if LH shocks can form and how we can distinguish them from their unmagnetized counterparts based on the sampled field data. We compare the time-evolution of the particle and field distributions in an unmagnetized plasma and in a magnetized plasma. The blast shell is created by letting a plasma with a high thermal pressure expand into one with a low thermal pressure. The plasmas in the high-pressure region and in the low-pressure region are both spatially uniform and at rest in their respective domains at the simulation's start. Our simulation setup thus differs from that in the related Ref. \cite{Forslund71}, which created the unmagnetized shock by letting a plasma beam collide with a reflecting wall and the magnetized shock with a magnetic pressure gradient. 

The low-pressure plasma consists of $N^{7+}$ ions and electrons with the same density and temperature that were considered in the previous section. We will refer to it as the ambient plasma. The high-pressure plasma has a density, which exceeds that of the ambient plasma by the factor 10. The electron temperature of the high-pressure plasma is 3 times higher than that of the ambient plasma, while the ion temperature is the same in both plasmas. The ambipolar electric field, which develops at the jump of the thermal pressure between the high-pressure plasma and the ambient plasma, forms a double layer \cite{Hershkowitz81,Schamel86} that lets the plasma expand in the form of a rarefaction wave \cite{Crow75,Mora05} until a shock forms. 

Our simulations will show that the actual shock forms in the ambient plasma well ahead of the expanding high-pressure plasma and far away from the location where the rarefaction wave was launched. The formation process should thus be independent of the idealized initial conditions.

The simulation box spans the interval -10 mm $<$ x $<$ 10 mm along the x-direction and this interval is subdivided into 4000 simulation cells. The high-pressure plasma is located in the interval -4 mm $<$ x $<$ 4 mm and it is surrounded by the ambient plasma. We use periodic boundary conditions that connect the ambient plasma at -10 mm with that at 10 mm and consider only the shock that forms in the half-space x $>$ 0. The simulation is stopped before the shock-accelerated ions reach the boundaries. 

We introduce an initial magnetic field $\mathbf{B} = B_0 \mathbf{z}$ with $B_0 = 0.85$ T in one simulation, while the plasma in the second one is unmagnetized. Both simulations use $6.4 \times 10^7$ computational particles (CPs) to represent the high-pressure electrons and ions, respectively. The ambient electrons are represented by a total of $1.6 \times 10^7$ CPs and the same holds also for the ambient ions. We normalize time to the inverse of the ion plasma frequency $\omega_{pi} = 1.55 \times 10^{10}$ rad $\mathrm{s}^{-1}$ of the ambient medium and one time unit thus corresponds to 65 ps. We compare the results of both one-dimensional simulations at the times $t_1= 1.8$, $t_2=7.4$, $t_3=17.5$ and $t_4=32.7$.
  
\begin{figure}[ht]
\includegraphics[width=\columnwidth]{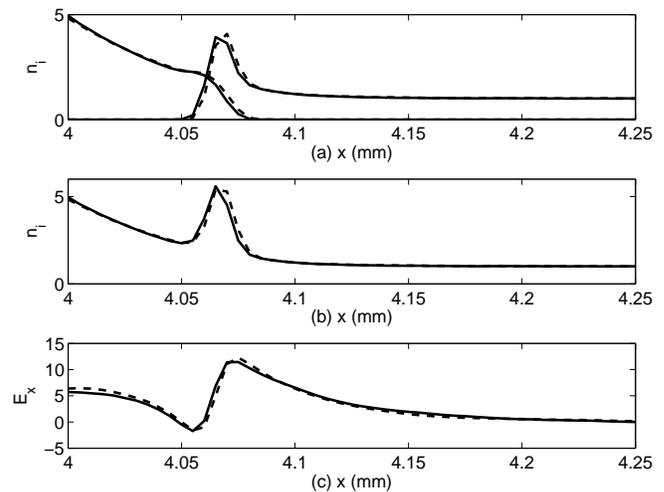}
\caption{Ion density normalized to $n_{i0}$ and electric field normalized to $10^7$ V/m at $t_1$ = 1.8: Panel (a) shows the density distributions of the magnetized and unmagnetized plasmas. The contributions of the high-pressure- and of the ambient plasma are displayed separately and the latter is located to the right. The cumulative ion distributions are shown in (b). Panel (c) shows the electric field distributions. Solid curves correspond to the magnetized plasma and dashed curves to the unmagnetized one.}\label{Density1dfirst}
\end{figure}

\textit{Time $t_1$ = 1.8}: The ion density distributions in both simulations and the electric fields are compared in Fig. \ref{Density1dfirst}. The ion density distributions in both simulations are practically identical; the magnetic field has not affected the ion expansion at this time. 
The matching distributions of the electric field $E_x(x,t)$ support this conclusion. An overlap layer, which is formed by the ions of both plasmas, is observed close to x = 4.07 mm. This overlap layer is the first stage of the shock formation \cite{Ahmed13}.

\begin{figure}[ht]
\includegraphics[width=\columnwidth]{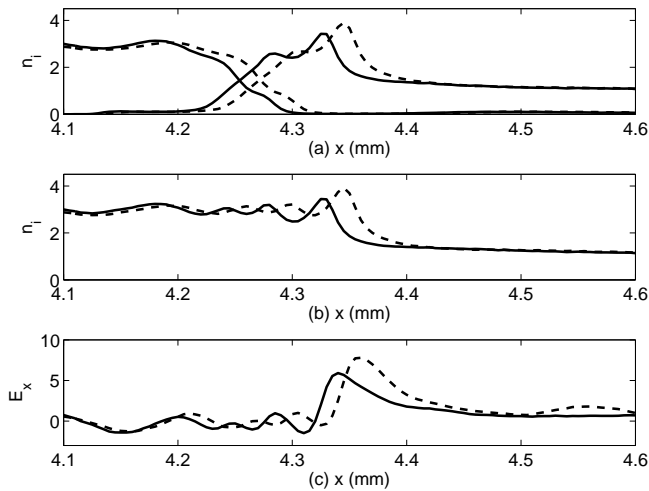}
\caption{Ion density normalized to $n_{i0}$ and electric field normalized to $10^7$ V/m at $t_2$ = 7.4: Panel (a) shows the density distributions of the magnetized and unmagnetized plasmas. The contributions of the high-pressure- and of the ambient plasma are displayed separately and the latter is located to the right. The cumulative ion distributions are shown in (b). Panel (c) shows the electric field distributions. Solid curves correspond to the magnetized plasma and dashed curves to the unmagnetized one.
}\label{Density1dsecond}
\end{figure}

\textit{Time $t_2$ = 7.4}: The ion density hump has spread out in space, forming a plateau between 4.1 mm and 4.25 mm in Fig. \ref{Density1dsecond}. 
A rarefaction wave with a density that decreases with increasing x is still present in the interval x $<$ 4.1 mm (not shown). The plateau is bound to the right by a shock, which is followed by the oscillations that are known to trail collisionless shocks \cite{Greenberg60}. The density peak is located at x $\approx$ 4.35 mm in the unmagnetized plasma and at x $\approx$ 4.33 mm in the magnetized plasma. The electric field distribution confirms that the blast shell has expanded farther in the unmagnetized plasma than in the magnetized one.
  
\begin{figure}[ht]
\includegraphics[width=\columnwidth]{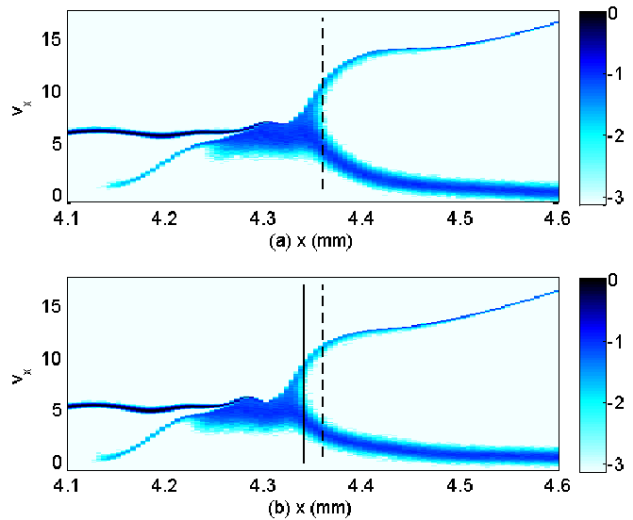}
\caption{Panel (a) shows the phase space density distribution $f_i(x,v_x)$ of the unmagnetized ions and panel (b) that of the magnetized ions at the time $t_2 = 7.4$. The color scale is 10-logarithmic, the densities are normalized to their peak value and velocities are normalized to $10^5$ m/s. The vertical dashed line shows the position x = 4.36 mm of the unmagnetized shock and the solid vertical line the position x = 4.34 mm of the magnetized shock.}\label{Phase1dsecond}
\end{figure}

Figure \ref{Phase1dsecond}(a) compares the phase space density distribution $f_i(x,v_x)$ of the ions in the magnetized plasma with that of the ions in the unmagnetized plasma at the time $t=t_2$.
The phase space density distributions are identical besides the lag of 20 $\mu$m of the shock in the magnetized plasma relative to that in the unmagnetized one. The simulation time $t_2$ corresponds to two percent of an ion gyroradius in the field $B_0$ since $\omega_{ci} / \omega_{pi} = 2.6 \times10^{-3}$. The ion's gyromotion is negligible at this time and the slowdown of the magnetized shock is not caused by the ion's rotation in the magnetic field. 

In Fig. \ref{Phase1dsecond} we find the ambient ions at $x>$ 4.4 mm and $v_x \approx 0$. A small fraction of the ions is reflected by the shock and they feed the shock-reflected ion beam at x $>$ 4.4 mm and $v_x \ge 10^6$ m/s. The remaining ions make it into the downstream region and form the hot population in the interval 4.25 mm $<$ x $<$ 4.35 mm and $4 \times 10^5 \mathrm{m/s} < v_x  < 6 \times 10^5 \mathrm{m/s}$. The high-pressure plasma's ions form the cool dense ion beam at $5 \times 10^5 \mathrm{m/s}$ in the interval x $<$ 4.25 mm. These ions do not mix with the hot downstream population and their density is negligibly small already at x = 4.3 mm (See Fig. \ref{Density1dsecond}(a)). The high-pressure ions thus serve as a piston but they do not mix with the ions close to the shock. The LH waves close to the shock should thus obey the wave dispersion relation of the ambient plasma shown in Fig. \ref{DispRel}. The speed, with which the high-density plasma is pushing the ambient plasma, is more than twice the phase speed of the LH wave estimated in Section 2 for a wavelength of 0.1 mm.

\begin{figure}[ht]
\includegraphics[width=\columnwidth]{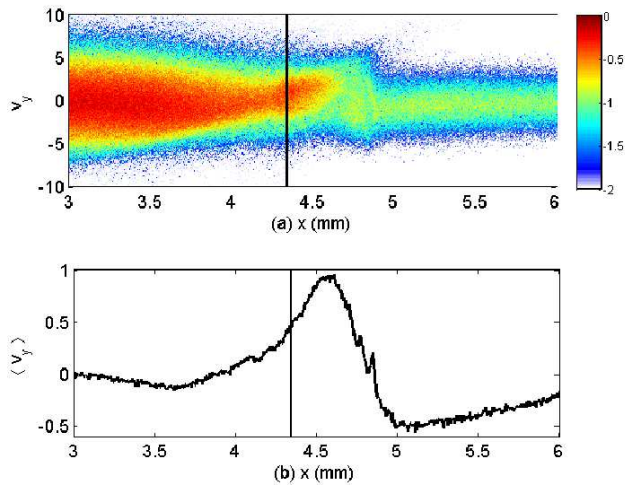}
\caption{The electron drift at $t_2$ = 7.4. Panel (a) shows the 10-logarithmic electron phase space density distribution $f_e (x,v_y)$ and panel (b) the electron drift speed $\langle v_y \rangle (x)$ expressed in units of $10^7$ m/s. The vertical line x = 4.34 mm is overplotted in both panels.}\label{Electrons50}
\end{figure}

Figure \ref{Density1dsecond}(c) evidences a strong electrostatic field $E_x$ in both simulations. The magnetic field with the strength $B_z$ = 0.85 T in the magnetized simulation will yield a $\mathbf{E} \times \mathbf{B}$-drift of the electrons relative to the ions. Figure \ref{Electrons50} reveals the magnitude of the electron drift. 
The electron phase space density distribution $f_e (x,v_y)$ shows a clear modulation close to the location x = 4.34 mm of the magnetized shock. The mean speed $\langle v_y \rangle (x) = \int v_y f_e (x,v_y) dv_y$ reveals that the drift speed is a sizeable fraction of $v_{te}\approx 1.9 \times 10^7$ m/s over a wide spatial interval. The peak drift speed is $\approx v_{te}/2$ or about $10^7$ m/s. 

The speed gain of the electrons at the shock exceeds that of the ions at the shock by the factor 20. The energy gain of the electrons is thus not negligible compared to that of the ions. The thermal pressure gradient of the dense plasma, which drives the shock, is the same in both simulations at this time and we may attribute the slower speed of the magnetized shock to this electron acceleration along the y-direction. The drastic change of the electron distribution at x $\approx$ 5 mm marks the front of the shock-reflected ion beam, where the ion density jump results in a jump of the electrostatic potential. 

\begin{figure}[ht]
\includegraphics[width=\columnwidth]{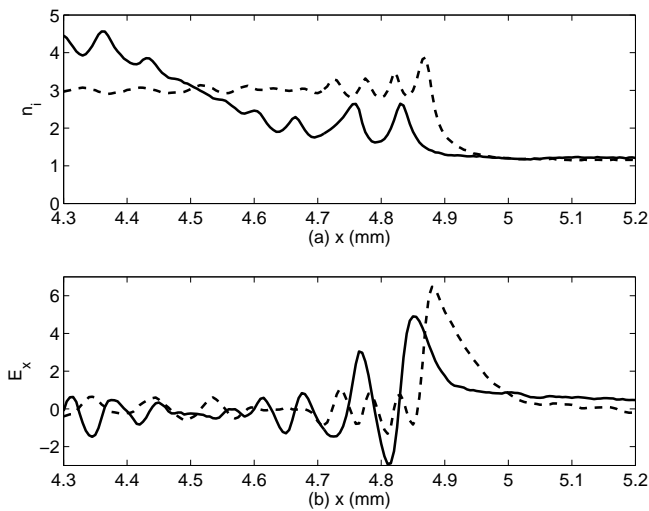}
\caption{Ion density normalized to $n_{i0}$ and electric field normalized to $10^7$ V/m at $t_3$ = 17.5: Panel (a) shows the cumulative distributions of the magnetized and unmagnetized plasmas. Panel (b) shows the electric field distributions. Solid curves correspond to the magnetized plasma and dashed curves to the unmagnetized one.
}\label{Density1dthird}
\end{figure}

Figure \ref{Density1dthird} reveals two important differences between both simulations at the time $t_3$ = 17.5. Firstly, the ion density in the magnetized simulation and in the region 4.6 mm $<$ x $<$ 4.9 mm is well below that in the simulation with $B_0 = 0$. The density oscillates around $3n_{i0}$ in the unmagnetized simulation and around $2n_{i0}$ in the magnetized one. The density oscillations in the magnetized simulation have a larger wavelength and amplitude than those behind the unmagnetized shock. The oscillations of the electrostatic potential will thus be much larger. 
The wavenumber, which corresponds to the wavelength of the oscillations $\approx$ 0.1 mm in the magnetized plasma, is indicated in Fig. \ref{DispRel} and we find only LH waves at low frequencies in this interval. 

The electron Bernstein modes and the upper-hybrid mode are fast electronic modes and such waves can not interact resonantly with ions that move at speeds that are much lower than $v_{te}$ \cite{Dieckmann00} and therefore these high-frequency waves are not destabilized by the ion beam.

\begin{figure}[ht]
\includegraphics[width=\columnwidth]{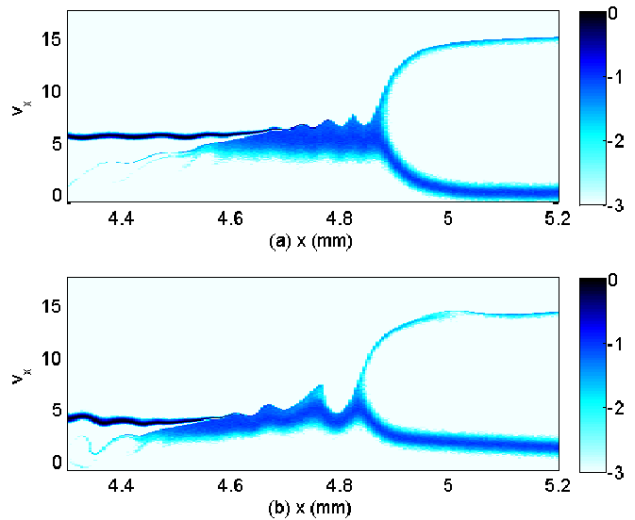}
\caption{Panel (a) shows the phase space density distribution $f_i(x,v_x)$ of the unmagnetized ions and panel (b) that of the magnetized ions at the time $t_3 = 17.5$. The color scale is 10-logarithmic, the densities are normalized to their peak value and velocities are normalized to $10^5$ m/s.}\label{Phase1dthird}
\end{figure}

Figure \ref{Phase1dthird} demonstrates that the trailing waves in both simulations are strong enough to visibly modulate the ion distribution. The large amplitude of the electrostatic oscillations affects in particular the downstream ion population of the magnetized structure. The amplitude of the velocity modulation at x $\approx$ 4.8 mm is larger than the thermal spread of the ions and it is close to that needed for the formation of ion phase space vortices \cite{Eliasson06}.

The magnetized structure and the electrostatic shock in Figs. \ref{Density1dthird} and \ref{Phase1dthird} show several important differences. A collisionless shock is characterized by a ramp with strong electric fields. The ramp of the unmagnetized shock is located in the interval 4.87 mm $<$ x $<$ 5 mm. The electric fields in the magnetized simulation are largest for 4.83 mm $<$ x $<$ 4.87 mm and naturally we would associate this interval with the ramp. 

The electric field in the magnetized simulation reaches out beyond x = 5 mm. If the magnetized structure were a shock, the upstream region with the weak electric field would be its foot. A foot is a feature of collisionless shocks. It is created by the ions, which have been reflected by the shock. A foot usually stretches out by about an ion gyroradius if the shock is perpendicular.  The gyroradius of the reflected ions with a speed $v_x \approx 10^6$ m/s and $B_0$ = 0.85 T will be 25 mm and the foot is thus still developing at this time.

Figure \ref{Phase1dthird}(b) shows that the ambient ions are already accelerated in the foot well before the ramp arrives. The ambient ions at the front of the ramp at x = 4.9 mm have reached a speed $\approx 3 \times 10^5$ m/s and the velocity gap between the ions of the downstream plasma and those of the upstream plasma has been reduced substantially. The plasma compression by the structure at $x\approx 4.83$ mm in Fig. \ref{Density1dthird}(a) is weaker than that observed for its unmagnetized counterpart, suggesting that the magnetized structure depicted in Fig. \ref{Phase1dthird}(b) at the same position is either a weak shock or a propagating nonlinear LH wave. In what follows we refer to it as nonlinear LH wave.

We want to determine the process that pre-accelerated the ambient ions. Any ion acceleration in a collisionless plasma must be tied to an electric field and this acceleration mechanism does apparently not work in the unmagnetized plasma. 

An upstream electric field can be driven by the current of the shock-reflected ions. A perpendicular magnetic field limits the electron's mobility in the ambient plasma and thus their ability to react to this field. Hence the consequences of the electric field will be different in both simulations. It is, however, not this electric field that pre-accelerates the ambient ions because such a field should reduce the net ionic current ahead of the shock. The acceleration of ambient ions towards positive x will, however, enhance it and we can rule out this process. 

\begin{figure}[ht]
\includegraphics[width=\columnwidth]{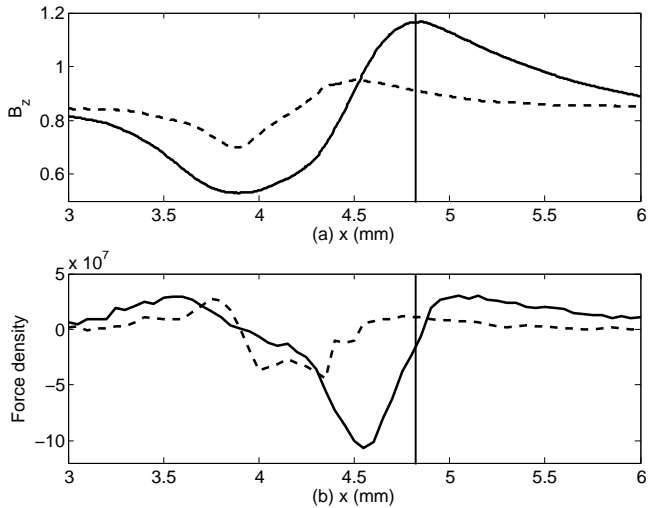}
\caption{The spatial distribution of magnetic $B_z(x)$ obtained from the simulation with the magnetized plasma is shown in panel (a) at the times $t_2$ =7.4 (dashed curve) and t=17.5 (solid curve). Panel (b) shows a moving average over 10 grid points of the associated force density $F_B=-{(2\mu_0)}^{-1}dB_z^2/dx$. The vertical line shows the location x=4.82 mm of the magnetized structure at $t_3$ = 17.5}\label{Magnetic}
\end{figure}

The $\mathbf{E}\times \mathbf{B}$ drift of the electrons in Fig. \ref{Electrons50} will modulate the magnetic field. Figure \ref{Magnetic}(a) compares the spatial distributions of $B_z$ at the times $t_2$ = 7.4 when the LH shock was strong and $t_3$ = 17.5 when it had transformed into a weaker shock or a nonlinear wave.
The expansion of the high-pressure plasma redistributes the uniform magnetic field $B_z$ = 0.85 T into a sine-like pulse that expands in space and time. The spatial extent of the pulse has grown by a factor 3 in space and the gradient has increased by about the same factor during the time interval $t_3-t_2$. The strong gradient gives rise to a gradient of the magnetic pressure $F_B = -{(2\mu_0)}^{-1}dB_z^2 / dx$. The force density changes sign at the nonlinear LH wave and it is of the order $\sim 10^7 N/m^3$ ahead of it in an x-interval with the size $\sim$ 1 mm. The force density accelerates ions with a mass density $m_i n_{i0} \sim 10^{-6} kg / m^3$, giving an acceleration $a_B \sim 10^{13}$ m/$s^2$. The acceleration time $t_a \sim 10^{-8}$ s of the ambient ions is given by the spatial extent of the acceleration zone $\sim$ 1 mm divided by the speed of the magnetized structure $\sim 10^5$ m/s. The product $a_B t_a \sim 10^5$ m/s is comparable to the speed gain of the ambient ions in Fig. \ref{Phase1dthird}(b) that led to the weakening of the shock.

The ion beam displayed in Fig. \ref{Phase1dthird}(b) at $x>4.9$ mm and $v_x > 7 \times 10^{5}$ m/s is a shock signature. The density compression in Fig. \ref{Density1dthird}(a), which is weak compared to that we observe at its unmagnetized counterpart, rules out that the structure in the simulation with $B_0 \neq 0$ is a shock that balances the ram pressure of the inflowing upstream medium solely with the downstream's thermal pressure. 

\begin{figure}[ht]
\includegraphics[width=\columnwidth]{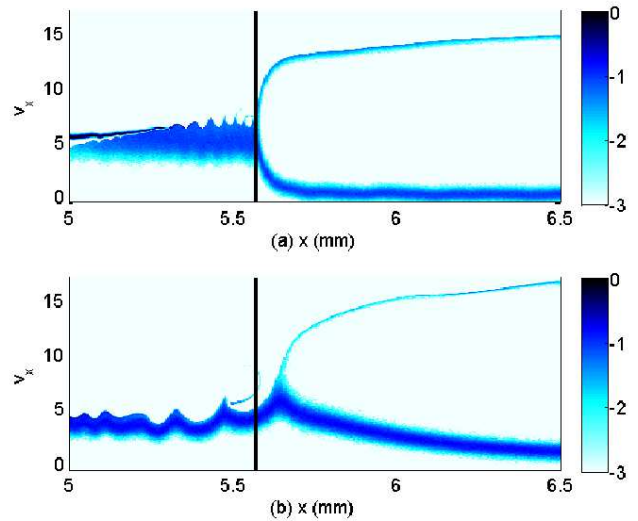}
\caption{Panel (a) shows the phase space density distribution $f_i(x,v_x)$ of the unmagnetized ions and panel (b) that of the magnetized ions at the time $t_4 = 32.7$. The color scale is 10-logarithmic, the densities are normalized to their peak value and velocities are normalized to $10^5$ m/s. The vertical line shows the position x = 5.57 mm of the unmagnetized shock.}\label{Phase1dfourth}
\end{figure}

Figure \ref{Phase1dfourth} compares the phase space density distributions of the ions at the time $t_4=32.7$. 
The unmagnetized shock remains qualitatively unchanged compared to its counterpart at the time $t_3$. The ions from the high-pressure plasma are visible at x $<$ 5.1 mm and $v_x \approx 5.5 \times 10^5$ m/s. They still act as a piston that pushes the ambient ions to increasing values of x. The shock is located at x = 5.57 mm and it separates the hot downstream ions from the cool ambient ions and the shock-reflected ion beam. 

The ion phase space density distribution in the magnetized simulation in Fig. \ref{Phase1dfourth}(b) resembles qualitatively that of the unmagnetized shock. We observe the pre-acceleration of ambient ions and a beam of ions that is reflected by the nonlinear LH wave. The nonlinear LH wave has overtaken the unmagnetized shock and it is located about 0.1 mm ahead of it. The reflected ion beam is still accelerated at x=6.5 mm; the magnetic pressure gradient force accelerates all ions far upstream of the nonlinear LH wave. We observe strong oscillations downstream of this structure, which are ion charge density oscillations and they are thus tied to the electrostatic LH waves.

Figure \ref{Phase1dfourth}(b) raises an important question. How can the magnetic pressure gradient force, which is a MHD force, sustain a nonlinear LH wave structure on a scale that is small compared to $r_{ci}$ and develop on a time-scale that is small compared to $\omega_{ci}^{-1}$: The ions have completed just about 1.4 \% of a gyro-orbit at the time $t_4$. 

The magnetic pressure gradient force also accelerates the electrons. The force density operates on electrons with the number density $7n_{i0} = 2.75 \times 10^{20} \textrm{m}^{-3}$ and the division of $F_B$ by $7n_{i0}$ gives us the force per electron. The electrons are accelerated along x and their current drives an electric field that counteracts the effects of $F_B$. The electron acceleration stops once a force balance in the x-direction is established and $F_B / 7 e n_{i0} = E_{bx}$, where $E_{bx}$ is the saturation field along x. 

\begin{figure}[ht]
\includegraphics[width=\columnwidth]{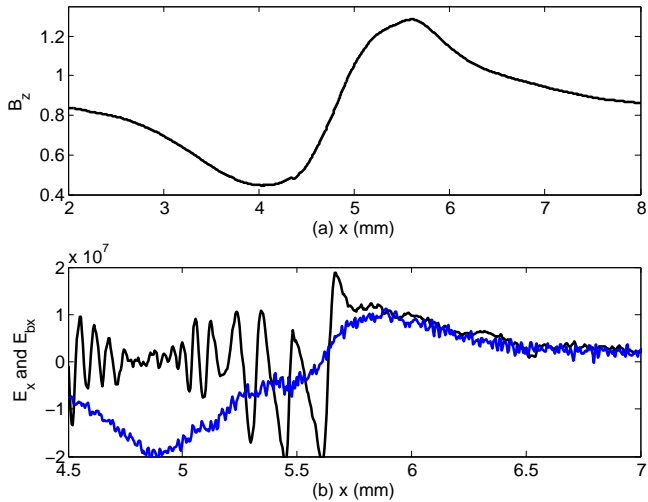}
\caption{The spatial distribution of magnetic $B_z(x)$ obtained from the simulation with the magnetized plasma is shown in panel (a) at the time $t_4$ = 32.7. Panel (b) compares the electric field $E_{bx}$ (blue curve with low-amplitude oscillations) that would balance the magnetic pressure gradient force to the electric field computed by the magnetized simulation.}\label{FinalForce}
\end{figure}

Figure \ref{FinalForce} shows $B_z (x)$ at the time $t_4 = 32.7$ and it compares $E_{bx}$ to the electric field, which is measured in the simulation. 
We observe that $E_{bx}$ matches the electric field $E_x (x)$ in the foot region of the nonlinear LH wave structure in the interval x $>$ 5.8 mm. 

The electric field driven by the magnetic pressure gradient force replaces the ambipolar electric field, which is tied to the thermal diffusion of electrons, with respect to sustaining the shock. For lower values of x, $E_x (x) \neq E_{bx}$. The mean value of $E_x (x)$ is in fact close to zero for 4.5 mm $<$ x $<$ 5.5 mm. The absence of an electric field implies that the magnetic pressure is balanced by the thermal pressure in this region, yielding a vanishing net force on the particles. Strong LH waves form in regions with a strong gradient of $E_{bx}$ and, thus, in an interval with a strong magnetic pressure gradient.  

One goal of our simulations is to determine how the field signature of an LH shock or a nonlinear LH wave differs from those of the well-researched electrostatic shocks in unmagnetized plasma. We have seen that an electrostatic shock approximately maintains its speed in the absence of a magnetic field and in one spatial dimension. Its magnetized counterpart is initially slower, but it overtakes the unmagnetized shock at a later simulation time. The structures in both simulations will also differ in their electric field distribution. A diagnostic technique that measures electromagnetic fields in laboratory plasmas should be able to distinguish both shocks. 

\begin{figure}[ht]
\includegraphics[width=\columnwidth]{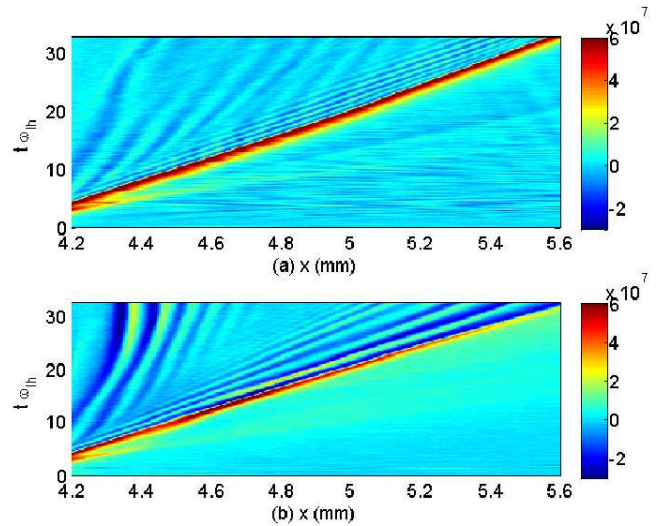}
\caption{The time-evolution of the electric field $E_x(x,t)$ in both simulations: Panel (a) and (b) show the field distribution in the unmagnetized plasma and in the magnetized plasma, respectively.}\label{Stack}
\end{figure}

A comprehensive overview of the electric field distribution is given by Fig. \ref{Stack}.
The electric field distribution of the shock in the unmagnetized plasma reveals a localized unipolar electric field pulse at all times. The pulse speed gradually decreases in time. The ion acoustic waves that trail the shock co-move with the shock and they keep their amplitude and wavelength constant. 

The unipolar pulse in the magnetized simulation has initially the same amplitude and width as its unmagnetized counterpart. The pulse amplitude, which characterizes the LH shock, decreases with time and it has almost vanished at $t_4=32.7$. The nonlinear steepening of the LH waves, which gave rise to the LH shock, is thus no longer sustained by the expanding plasma, substantiating that the LH shock is weakening or vanishing. 

We can not exclude that the LH shock will eventually reform. However, the time-scale of a cyclic reformation of fast magnetized shocks is of the order of the inverse ion gyrofrequency \cite{Lembege92}, exceeding by far our simulation time. Even if the reformation period is shorter for LH shocks than for their faster counterparts the instability between the two ion beams in the upstream would have to be taken into account. We also note that the piling up of the magnetic field ahead of the LH shock (See Fig. \ref{FinalForce}) may eventually result in a dynamic confinement of the expanding blast shell by the magnetic pressure in the ambient plasma. We leave a study of that long-term evolution to future work.

The waves that trail the shock increase their wavelength in time and they keep the electric amplitude unchanged. The electrostatic potential, which is associated with these oscillations, thus increases in time. The difference between the distribution of the electrostatic potential downstream of an unmagnetized and of a magnetized shock should be detectable. The same may hold for the different velocities of both shocks.

We must verify that the results obtained from the 1D PIC simulations are valid also in a more realistic two-dimensional geometry. Electrostatic shocks in more than one dimension are eventually destroyed by instabilities between the ambient ions and the shock-reflected ones. Their evolution is well-documented \citep{Karimabadi91,Kato10,Dieckmann14} and we will not consider further the unmagnetized case.

Drift instabilities can develop in the magnetized plasma due to the substantial $\mathbf{E}\times \mathbf{B}$-drift of the electrons with respect to the ions. The LH and electron cyclotron drift instabilities will drive waves with wavevectors that are aligned with the electron drift speed \cite{Brackbill84,Daughton04,Dieckmann14}. These drift instabilities are thus suppressed by the one-dimensional simulation geometry. 

We thus perform a two-dimensional PIC simulation with plasma parameters and with a box size along the x-direction that are identical to their counterparts in the magnetized simulation, which we have discussed above. We resolve the y-direction by 400 grid cells. The length of the box along y is 2 mm and the grid cell size along y is the same as the one along x. The number of CPs per cell is reduced by the factor 50 compared to that in the one-dimensional simulation.  

\begin{figure}[ht]
\includegraphics[width=\columnwidth]{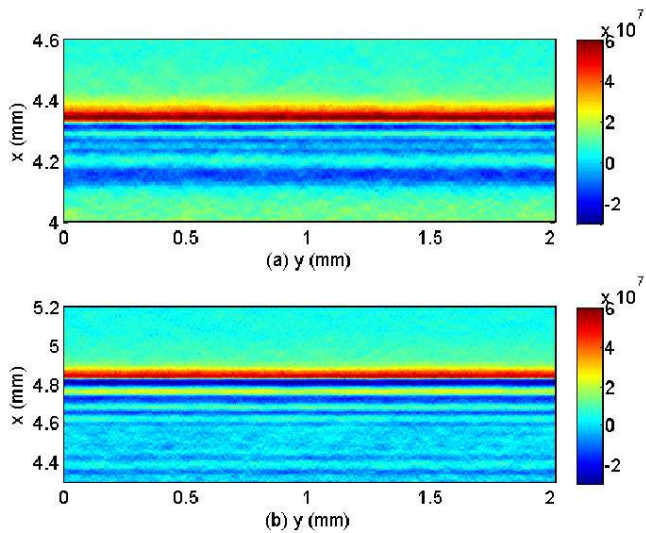}
\caption{The electric field $E_x(x,t)$ computed by the 2D PIC simulation: Panel (a) and (b) show the field distribution for the time $t_2$ = 7.4 and $t_3$ = 17.5, respectively.}\label{Electric2d}
\end{figure}

The electric field distributions $E_x(x,y,t_2)$ and $E_x(x,y,t_3)$ are displayed in Fig. \ref{Electric2d}. 
They are similar to those in the Figs. \ref{Density1dsecond}(c) and \ref{Density1dthird}(b). The electric field shows planar structures that are aligned with the y-axis and the dynamics of the plasma is thus one-dimensional. If a drift instability develops, then the resulting wave fields are too weak to be detectable and to affect the plasma dynamics. We do not show the electric field along the y-direction, because it consists solely of noise. 

\section{Discussion}

We have compared the expansion of a high-pressure plasma into a dilute ambient plasma with and without a perpendicular background magnetic field. The plasma parameters were comparable to those we find in laser-generated plasma.  We found strong electric field pulses in both simulations, which did correspond to the ambipolar electric field across a sharp plasma density change. These pulses expanded into the ambient plasma and accelerated it. The density of the ambient plasma was compressed by a factor 3 when it crossed the pulse and a fraction of the ambient plasma was reflected back upstream.

The pulse propagated in the simulation with no magnetic field into the ambient medium at the practically constant speed $\approx 2.5c_s$, where $c_s$ is the ion acoustic speed in the latter. The electric field pulse and the density change were thus an electrostatic shock. 

The introduction of the perpendicular magnetic field in the second simulation suppressed the ion acoustic wave. The lower-hybrid (LH) wave branch emerged and its phase speed at large wavenumbers was comparable to the value of $c_s$ in the unmagnetized plasma for our plasma parameters. The electric field pulse and the density jump in the simulation with the magnetized plasma moved at a speed above the phase speed of LH waves at large wavenumbers and the pulse corresponded to a LH wave shock. The structure of the LH shock resembled that of the electrostatic shock in the unmagnetized plasma apart from is slightly lower expansion speed. We have attributed the lower speed to the additional resistance imposed on the shock by the $\mathbf{E}\times \mathbf{B}$-drift of the electrons. 

The gradient of the magnetic pressure gave rise to a pre-acceleration of the ambient ions and the relative speed between the LH shock and these upstream ions decreased to a value below the phase speed. The LH shock changed into what appeared to be a nonlinear LH wave, which balanced the ram pressure of the inflowing ambient medium with the magnetic pressure and only to a lesser degree with the thermal pressure of the downstream medium. 

The increasing magnetic pressure in the ambient plasma suggests that eventually we may obtain a magnetosonic shock that separates the fast-moving and weakly magnetized (upstream) blast shell plasma from a slow-moving and strongly magnetized ambient (downstream) plasma. The electrostatic thermal-pressure gradient driven LH shock that separated the (downstream) blast shell plasma from the (upstream) ambient plasma would thus be only be a transient structure and its main effect would be to mediate the development of a magnetohydrodynamic shock.

A two-dimensional PIC simulation demonstrated that at least during the initial expansion phase the plasma dynamics remained one-dimensional close to LH shock. More specifically, the electron $\mathbf{E} \times \mathbf{B}$-drift current was not strong enough to drive LH wave turbulence close to the shock.

The LH wave has been invoked as a means to accelerate ions in the foreshock regions of magnetized shocks \cite{McClements97}. Boundary layers in a magnetized plasma that separate different ion populations have previously been found in hybrid simulations \cite{Chapman87,Chapman89,Niemann13}, which examined the demagnetization of an ambient plasma by an expanding plasma plume. Given the right plasma conditions, such boundary layers could steepen into an LH shock. 

The LH shock is trailed by LH waves with a larger electric field amplitude and wavelength than its unmagnetized counterpart. It should be possible to distinguish in laboratory experiments like the one performed in Ref. \cite{Niemann13} LH shocks from unmagnetized shocks based on the potential distribution that is trailing the shock. 

\textbf{Acknowledgements:} G. Sarri wishes to acknowledge EPSRC (Grant number: EP/N022696/1. The simulation was performed on resources provided by the Swedish National Infrastructure for Computing (SNIC) at HPC2N (Ume\aa). We thank the referee for the simple yet accurate linear dispersion relation for the LH wave.

\end{document}